\documentclass{llncs}
\usepackage{graphicx}

\usepackage[sorting=none, maxnames=50]{biblatex}
\usepackage{amsmath}
\usepackage{color}
\usepackage{url}
\usepackage{booktabs} 
\usepackage{amsfonts}
\DeclareGraphicsExtensions{.pdf,.png,.jpg}
\usepackage[english]{babel}
\usepackage[table,xcdraw]{xcolor}
\usepackage{listings}
\usepackage{xcolor}
\definecolor{codegreen}{rgb}{0,0.6,0}
\definecolor{codegray}{rgb}{0.5,0.5,0.5}
\definecolor{codepurple}{rgb}{0.58,0,0.82}
\definecolor{backcolour}{rgb}{0.95,0.95,0.92}
\lstdefinestyle{mystyle}{
    backgroundcolor=\color{backcolour},   
    commentstyle=\color{codegreen},
    keywordstyle=\color{magenta},
    numberstyle=\tiny\color{codegray},
    stringstyle=\color{codepurple},
    basicstyle=\ttfamily\footnotesize,
    breakatwhitespace=false,         
    breaklines=true,                 
    captionpos=b,                    
    keepspaces=true,                 
    numbers=left,                    
    numbersep=5pt,                  
    showspaces=false,                
    showstringspaces=false,
    showtabs=false,                  
    tabsize=2
}
\newcommand\YAMLcolonstyle{\color{red}\mdseries}
\newcommand\YAMLkeystyle{\color{black}\bfseries}
\newcommand\YAMLvaluestyle{\color{blue}\mdseries}
\makeatletter
\newcommand\language@yaml{yaml}
\expandafter\expandafter\expandafter\lstdefinelanguage
\expandafter{\language@yaml}
{
  keywords={true,false,null,y,n},
  keywordstyle=\color{darkgray}\bfseries,
  basicstyle=\YAMLkeystyle,                                 
  sensitive=false,
  comment=[l]{\#},
  morecomment=[s]{/*}{*/},
  commentstyle=\color{purple}\ttfamily,
  stringstyle=\YAMLvaluestyle\ttfamily,
  moredelim=[l][\color{orange}]{\&},
  moredelim=[l][\color{magenta}]{*},
  moredelim=**[il][\YAMLcolonstyle{:}\YAMLvaluestyle]{:},   
  morestring=[b]',
  morestring=[b]",
  literate =    {---}{{\ProcessThreeDashes}}3
                {>}{{\textcolor{red}\textgreater}}1     
                {|}{{\textcolor{red}\textbar}}1 
                {\ -\ }{{\mdseries\ -\ }}3,
}
\lst@AddToHook{EveryLine}{\ifx\lst@language\language@yaml\YAMLkeystyle\fi}
\makeatother
\newcommand\ProcessThreeDashes{\llap{\color{cyan}\mdseries-{-}-}}

\DeclareUnicodeCharacter{21B2}{\lefturn}
\begin{document}

\frontmatter          
\pagestyle{empty}  

\title{Implementing a scalable and elastic computing environment based on Cloud Containers}

\author{Damiano Perri\inst{1,2} $^{ORCID: 0000-0001-6815-6659}$ 
\\Marco Simonetti\inst{1,2} $^{ORCID: 0000-0003-2923-5519}$ 
\\Sergio Tasso\inst{2} $^{ORCID: 0000-0001-9174-9065}$  
\\Federico Ragni\inst{3} and
\\Osvaldo Gervasi\inst{2} $^{ORCID: 0000-0003-4327-520X}$ 
}
\institute{
University of Florence, Dept. of Mathematics and Computer Science, Florence, Italy \and University of Perugia, Dept. of Mathematics and Computer Science, Perugia, Italy \and University of Perugia, Information Systems Division, Perugia, Italy
}

\titlerunning{Implementing a scalable and elastic computing environment based on Cloud Containers} 
\authorrunning{D. Perri, M.  Simonetti, F. Ragni, S. Tasso and O. Gervasi} 

\maketitle

\begin{abstract}
In this article we look at the potential of cloud containers and we provide some guidelines for companies and organisations that are starting to look at how to migrate their legacy infrastructure to something modern, reliable and scalable.
We propose an architecture that has an excellent relationship between the cost of implementation and the benefits it can bring, based on the "Pilot Light" topology. 
The services are reconfigured inside small docker containers and the workload is balanced using load balancers that allow horizontal autoscaling techniques to be exploited in the future.
By generating additional containers and utilizing the possibilities given by load balancers, companies and network systems experts may model and calibrate infrastructures based on the projected number of users.
Containers offer the opportunity to expand the infrastructure and increase processing capacity in a very short time.
The proposed approach results in an easily maintainable and fault-tolerant system that could help and simplify the work in particular of small and medium-sized organisations.

\end{abstract}


\section{Introduction}
The cloud is one of the biggest technological innovations of recent years \cite{10.1145/1364782.1364786,antonopoulos2010cloud}.
In fact, it enables the creation of complex, reliable and available technological infrastructures that provide services of various kinds:
from calculation services to storage ones, to servers for the contents' distribution via web pages.
There are various forms of cloud: for example we can use a public cloud or set up a private cloud \cite{basmadjian2012cloud,doelitzscher2011private,li2011comparing}.
In the case of a public cloud, hardware provided by third-party companies is used, such as Amazon AWS, Microsoft Azure, Google Cloud, and so on.
Through small initial investments, only what is actually used is paid for, and the hardware which is made available to the users, will be maintained by the companies.

In contrast to the public cloud, there is the private cloud. 
This term refers to infrastructures that are typically corporate, providing virtualised and reduntant hardware and highly reliable services for employees, associated staff, or a external users who consume the content produced by the company.
In this case, the company will have the responsibility of maintaining and configuring the hardware needed to set up and build the infrastructure.
This initial disadvantage, however, is counterbalanced by a huge strategic advantage: 
the data (either sensitive either ultra-sensitive data) is completely entrusted to the management of the company itself and does not have to transit to third-party servers, so this solution may be preferable for certain applications \cite{ren2012security}.
The private cloud is a type of development architecture in which the computational resources are reserved and dedicated for the organisation managing the system. 

In the public cloud, on the other hand, the services offered by a provider use pools of machines that also accommodate other users. 
There is a third alternative, called hybrid cloud, which represents a combination of the two architectural strategies \cite{li2014hybrid}.
In the hybrid cloud, part of the services are given by a third-party provider, and some subsystems are allocated within the corporate platform.
For example, it is reasonable to imagine a situation where the provider is demanded to maintain a certain number of virtual machines and a certain number of disks for data storage while a second pool of hard disks containing encrypted sensitive data or databases for permanent information storage, is located inside the private organisation.

The Covid-19 pandemic has highlighted a feature common to many countries around the world: many organisations and companies have inadequate infrastructure to meet the pressing needs arising as a result of the digitisation processes that have become extremely urgent.
Our intent is to provide a model that can speed up the technological evolution of companies and organisations or improve their current computing infrastructure.
The solution we propose, supported by practical experimentation, allows a transition from an obsolete infrastructure of an SME (small and medium-sized enterprise), representing a typical case, to a system based on a cluster whose services are deployed by docker containers, and delivered through a cluster of nodes configured according to the best practices of an High Availability (HA) approach.
The idea sprang from a report of the Italian National Institute of Statistics (STAT), which examined active companies with at least 10 employees, finding out that the Information Technology (IT) infrastructure are still inadequate in many cases to their needs.\footnote{\url{http://dati.istat.it/Index.aspx?DataSetCode=DCSP_ICT}}

\section{Related works}
The cloud is being studied and analysed by many researchers around the world because of its strong capabilities.
In the course of this decade, it is expected that the cost-benefit ratio will continue to increase and there will be new patterns of developments focused on the Internet of Things (IOT) \cite{10.1145/3241737}, and other emerging technologies.
The impacts that this type of architecture have on the environment are also studied.
The data centres which are required by the Cloud, consume large amounts of energy and it is necessary to make accurate estimates of the pollution that will be produced in the next few years \cite{10.1145/3241038}.
The cloud can also be used to make sites and content available for use in the world of education \cite{10.1007/978-3-030-58820-5_56,10.1007/978-3-030-58820-5_57}.
A number of researchers are tackling extremely topical and interesting subjects, e.g. techniques for implementing and exploiting Function as a Service (FaaS) \cite{10.14778/3407790.3407836}.
FaaS are serverless systems, where programmers can insert their own snippets of code (Python, PHP, etc.) and call them up via APIs.
The result is that virtual machines automatically execute the code, without the necessity of setting up a server.
The cloud can also be used to do complex computational calculations that require expensive GPUs and CPUs to perform machine learning calculations \cite{bocca2019,perri2020binary,cancroPelle,cnnDamianoDividiti2019}.

Docker containers are lightweight cloud technologies that are dominating among IT solutions because they allow applications to be released faster and more efficiently than in the case of virtual machines. The adoption of Docker containers in dynamic, heterogenous environments and the ability to deploy and effectively manage containers across multiple clouds and data centers has made these technologies dominant and fundamental \cite{cloudContainers}. The improvements in terms of increased performance and reduced overhead have made the cloud container approach indispensable for building cloud environments that keep pace with the demands emerging from various application domains \cite{ containerOverhead,containerSQL}.

\section{Towards scalable and reliable services}
Legacy service delivery architectures are based on a single, centralised server.
The main disadvantage of this solution is its low maintainability and the lack of fault tolerance.
If a hardware problem or a hacker attack occur, service delivery may be compromised.
Monolithic architectures also suffer from another disadvantage: they cannot effectively scale in case of peak demands.

Other types of architectures support scaling.
Scaling can be of two types: vertical or horizontal \cite{gong2010press,vaquero2011dynamically,mao2010cloud}.
Vertical scaling is defined as an operation that increases the  machine hardware resources, for example, rising the number of vCPUs or the amount of GiB of RAM.
Horizontal scaling is defined as the operation that creates replicas of the server (node) that provides the service.
The new node is identical to the original one and will help it to respond to client requests.
In the case of monolithic architectures, horizontal scaling is not possible.

Vertical scaling, on the other hand, involves shutting down the machine and upgrading the hardware.
This is unsuitable and not acceptable by modern standards.
The problem of fault tolerance must also be taken into account.

\subsection{Disaster recovery}
Legacy architectures often do not have particularly complex plans for fault management or data loss \cite{alhazmi2012assessing,alhazmi2013evaluating}.
Disaster recovery (DR) is the procedure implemented to restore the functionality of a system, suffering a disaster, a damage: 
for example, the loss of system data, the loss of user data, the compromise of security following a hacker attack or hardware damage due to a natural disaster.
When planning internal disaster recovery policies, three objectives must be defined: 
the Service Level Agreement (SLA), the Recovery Time Objective (RTO) and the Recovery Point Objective (RPO) \cite{chang2015towards,khoshkholghi2014disaster}.

The SLA is a percentage value that indicates the minimum uptime that the system guarantees during the year: for example, if a system has an SLA value of 99\%, it means that in one year it could be unavailable for 3d 15h 39m 29s.
An SLA of 99.99\% instead could accept an annual downtime of 8h 45m 56s.
The higher the desired SLA value, the higher the costs to build an architecture to meet our demand.

The RTO is the maximum acceptable time between the interruption of the service and the moment when it is restored.
In the case of legacy architectures the RTO is generally about 48 hours; for example, in case of hardware problems, spare parts must be found, repairs or replacements must be made, and the system must be restored.

The RPO indicates the maximum acceptable time between the last data backup and time data loss because of a disaster. 
This indicates how much data time, in terms of hours, we can accept to lose.
Legacy architectures often rely only on RAID 1 storage systems (mirroring) and do not perform daily incremental and offsite backups.
Unlike the SLA, we have that the lesser RTO and RPO time is required by the system, the higher it is the cost to achieve this requirement.
\begin{figure}[ht]
    \centering
    \includegraphics[width=\linewidth]{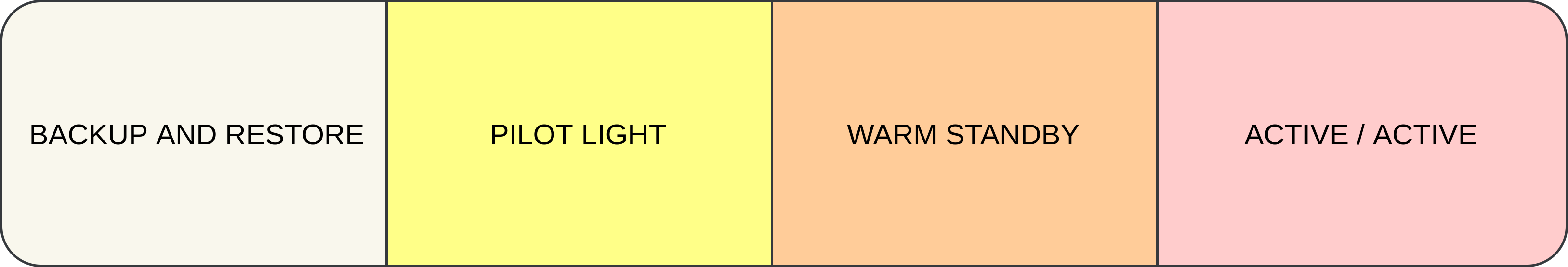}
    \caption{Disaster recovery plans}
    \label{fig:disasterRecovery}
\end{figure}
The types of disaster recovery plans\cite{hamadah2019cloud} that can be implemented are summarised in Figure \ref{fig:disasterRecovery}.

Generally, legacy architectures implement a "Backup and Restore" type of architecture.
This means that in case of critical problems, the only thing that can be done is to shut down the machine (if it is still working), restore the backup on the repaired machine or on a new machine, and restart the system with the data updated to the last backup.
The "backup and restore" mode has an RPO in hours and an RTO in the 24 to 48 hour range.

"Pilot Light" mode creates hourly or daily backups and maintains a complete copy of the architecture in a separate place, away from the system.
In the event of a disaster, IT technicians will initiate backup and services will be restored.
The "pilot light" mode guarantees an RPO in minutes and an RTO in hours.

Alternatively, there is the "Warm Standby" mode.
In this mode, a backup system is replicated on a different location with respect to the main system and is synchronised in near real-time.
This mode provides for a synchronisation in seconds (RPO) and an RTO time in minutes.

On the other hand, the "Active / Active" mode is the last possible architecture according to the diagram shown, and it is the safest and the most expensive from the point of view of service availability.
It requires that two or more service delivery systems are always synchronised and active at the same time.
A load balancing service is required to sort requests between the two systems.
For example, an active / active system of the type 70 / 30 could be configured. 
A 70\% of the requests will go to the main system and a 30\% of the requests will go to the secondary backup system.

If the main system experiences a problem, the secondary backup system will be active and the user will not experience any particular problem.
This type of architecture provides for an RPO around milliseconds' time, or even 0, and an RTO potentially equal to 0.
It should be noted that, as we move towards the right-hand side of the Figure \ref{fig:disasterRecovery}, the protection and ability of the system to resist faults increases, so too the costs of implementing the architecture.

There are several problems with Legacy architectures: first, we can be exposed to ransomware hacking attacks that can irretrievably destroy our data store; second, there could be a software problem or a badly crafted query by a software developer that would irrevocably wipe out the database; therefore, other kinds of software problems can exist: minor portability issues among operating systems.

An example of legacy architecture with no virtual machines is shown in the Figure \ref{fig:legacy}A.

In past years, it was in fact common practice to install a Linux distribution and directly configure the application servers at the operating system level.
A further problem is related to system updates.
If, for example, it is necessary to update the PHP server in order to be able to use the new version and the latest security improvements, this might be difficult and problematic in the legacy case.
Furthermore, it would be necessary to assess how much downtime we can accept in order to carry out the update; but there would be no guarantee that a rollback could be carried out quickly in the event of incompatibility problems.

\subsection{Legacy with Virtual Machines}
\begin{figure}[ht]
    \centering
    \includegraphics[width=\linewidth]{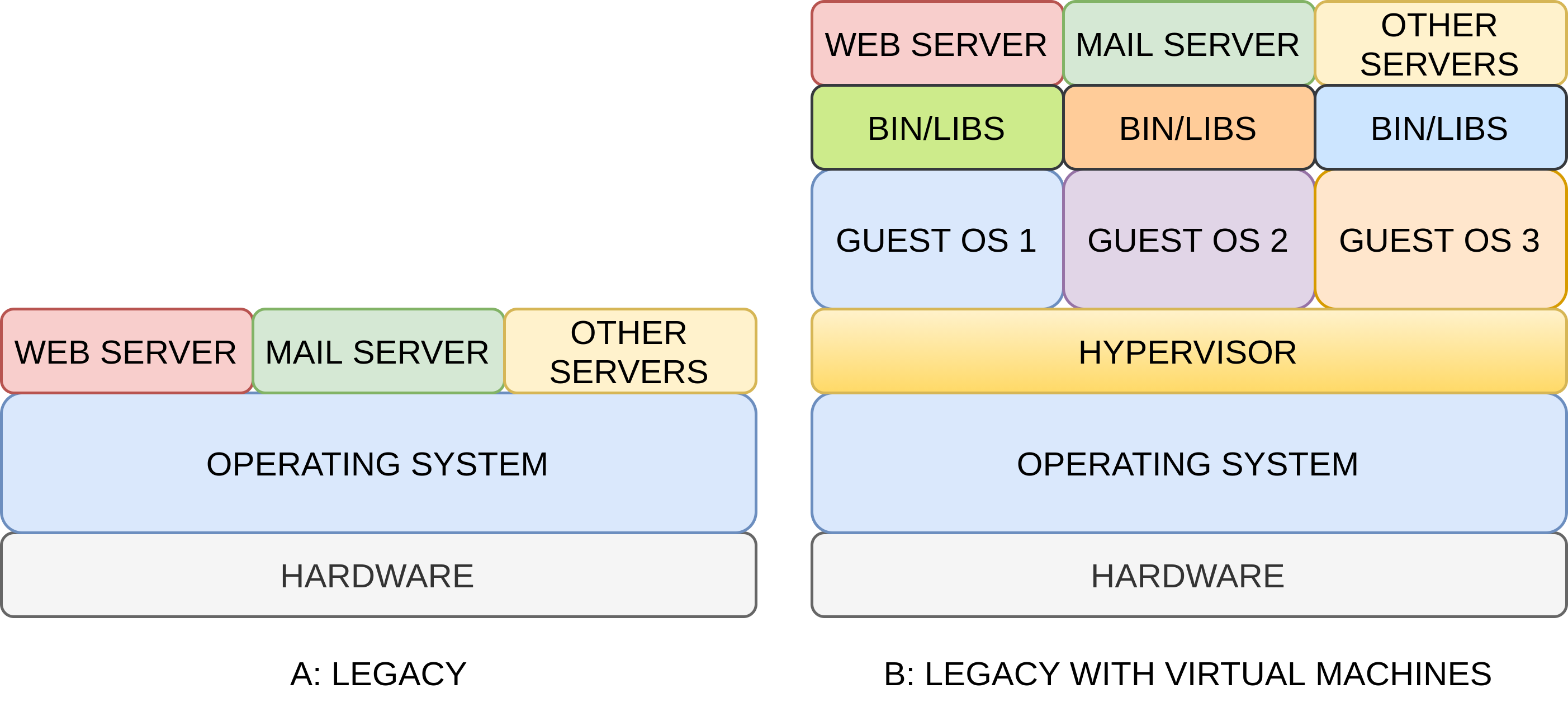}
    \caption{Legacy Architecture}
    \label{fig:legacy}
\end{figure}

A slightly better solution, falling however under legacy configurations, is to use several Virtual Machines (VMs), one for each service-providing application.
An example is shown in the Figure \ref{fig:legacy}B.
In the second scenario, using Virtual Machines, programs may be moved from one system to another, because virtual machines are kept in one or more files readily transferred across devices.
The side effect is that they have a significant hardware impact.
Since a complete system needs to be executed, a lot of hardware (especially RAM) resources are used even when not required by the applications.
In addition, further clock cycles are used to perform and maintain the secondary operating system activities, which also waste energy.
The installation of the operating system also requires that virtual machines occupy a considerable amount of disk space, each time that a new one is generated.
There are consequently two benefits of this architecture: the ability to make backups quickly, for example by duplicating the Virtual Machine-related files and the independence from the hardware subsystem and operating system.

\section{The proposed architecture}
In this section we describe the main techniques we used to implement a modern, reliable and highly available system (HA).
The technology behind the proposed architecture is based on the use of Docker containers, which have the enormous advantage of being much leaner and more efficient than a virtual machine. 
Containers also make it possible to isolate an application at the highest level, making it a completely separate entity.

\begin{figure}[ht]
    \centering
    \includegraphics[width=0.65\linewidth]{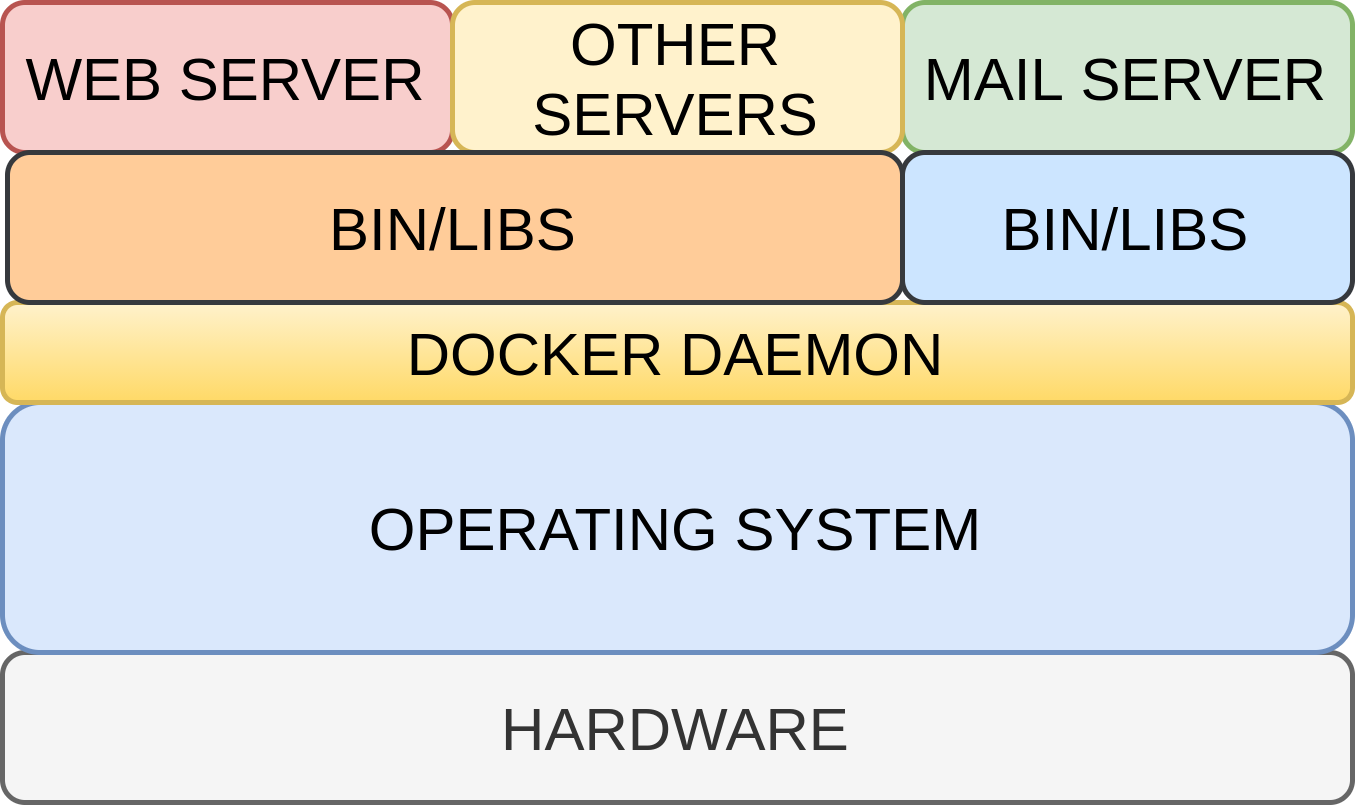}
    \caption{Docker architecture}
    \label{fig:docker}
\end{figure}

The first step is to design and configure the services and applications using Docker containers, as described in Subsection \ref{sub:docker} \cite{boettiger2015introduction,turnbull2014docker}.
Docker also guarantees high security thanks to its container architecture with separate storage spaces and access permissions \cite{combe2016docker}.
In the Subsection \ref{sub:DB} we explain the way we can set up the databases in a Master/Slave configuration.

\begin{figure}[ht]
    \centering
    \includegraphics[width=\linewidth]{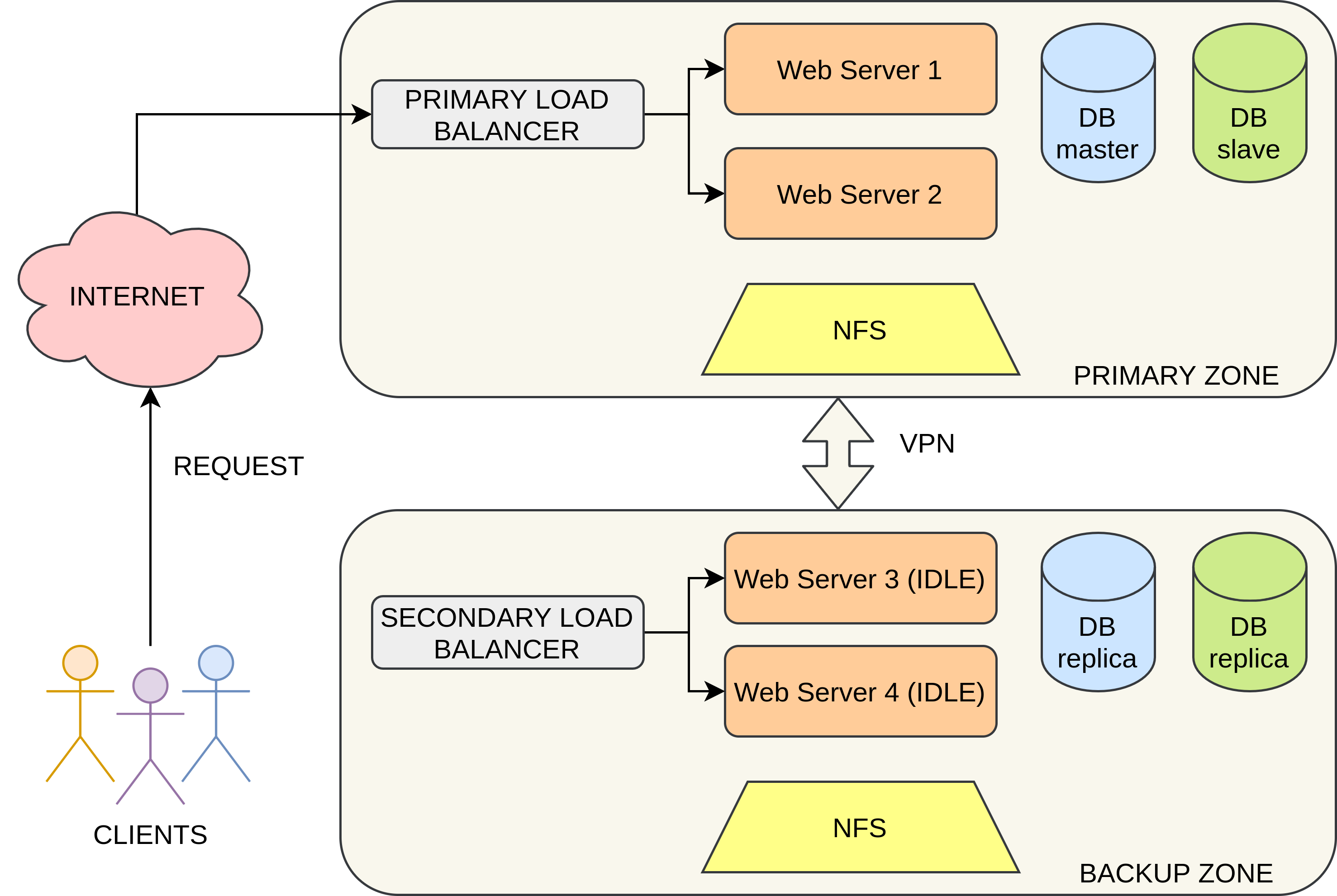}
    \caption{Pilot light architecture}
    \label{fig:proposedArchi}
\end{figure}

Once the services are up and running inside the containers, a distributed, redundant network environment must be prepared using load balancers, as described in Section \ref{sub:LB}.
A networked, redundant and available file system must then be set up as described in Section \ref{sub:NFS}.
The proposed architecture is shown in the Figure \ref{fig:proposedArchi}.
As it is shown, two availability zones are configured according to the Pilot Light scheme.
The two zones are connected to each other via a VPN connection.
The primary zone allows horizontal scaling thanks to the use of the load balancer and small servers inside docker containers.
The secondary zone is on IDLE state and is kept at minimum power and CPU consumption; services are configured but not active.
Data is copied between the two zones in an automated way.
In case there is a problem and zone A fails, the backup zone will take its place.
\subsection{Docker}\label{sub:docker}
Docker allows to build an architecture as shown in the Figure \ref{fig:docker}.
Each service, such as the Web Server or the Mail Server, is encapsulated within its own container.
Containers are defined by "yml" files, and an example of them is given in the code shown in Listing \ref{webyml}, where a Web Server Apache container is defined, exposing the HTTP and HTTPS ports.
Each "yml" file may contain the definition of one or more containers.
Containers are extremely light from a computational point of view and do not instantiate a real operating system; the applications that run inside them only need to allocate the libraries and binaries necessary for the application to work.
Containers are stateless by definition, which means they have no true running state.
To ensure data consistency, we should mount the folders of the filesystems where we want to execute input/output operations in a permanent way within them.

\begin{lstlisting}[language=yaml,frame=single,caption=Web Server without Load Balancer,label=webyml,float=h]
version: '3'
services:
  web:
    image: apache
    container_name: apache_web
    restart: always
    ports:
      - "80:80"
      - "443:443"
    volumes:
      - "/home/user/myWebsite:/var/www/html/"
    deploy:
      resources:
        limits:
          cpus: '4.0'
          memory: 2048M
        reservations:
          memory: 1768M
\end{lstlisting}

\subsection{The load balancing service}\label{sub:LB}
Load balancers are fundamental to the implementation of an HA architecture.
They are customizable and can be adapted to a wide range of applications.
The standard task of a load balancer is to distribute incoming requests to a pool of worker nodes.
The nodes will process the requested information and if necessary provide the output to the users.
The requests load can be managed in two different ways: balanced and unbalanced.

In the balanced mode each node receives a quantity of requests equal to 1/N, with N equal to the number of nodes.
This type of balancing can be implemented in the case of an Active/Active architecture.

In unbalanced mode, percentage values can be defined to indicate the amount of requests each node will receive. For example, a 60/40 configuration allows 60\% of requests to be sent to Node1 and 40\% of requests to Node2.

Moreover, there are various request scheduling algorithms \cite{prasetijo2016performance,pramono2018round}.
The first algorithm is called round-robin.
Requests are sorted cyclically across nodes using the round-robin algorithm. This method ensures that nodes receive an equal amount of requests regardless of their CPU use or complexity.
A second algorithm is Least Outstanding Requests (LOR) which sorts requests across nodes trying to balance the number of "unprocessed requests".
In our architecture, two load balancers are configured.
The first is HAproxy placed in a docker container.
HAproxy only exposes ports 80 HTTP and 443 HTTPS  \cite{prasetijo2016performance}.
Its role is to obtain requests from clients and sort them within the primary zone, evenly distributing the workload across nodes.
So, The file \texttt{haporxy.cfg} must be configured.
\begin{lstlisting}[language=yaml,frame=single,caption="Definition of the haproxy",label=haproxy_yml,float=h]
version: '3'
services:
 haproxy:
  image: haproxy:2.3.5
  hostname: haproxy
  ports:
    - 80:80
    - 443:443
  volumes:
    - /myHaproxy.cfg:/usr/local/etc/haproxy/haproxy.cfg
  deploy:
    placement:
     constraints: [node.role == worker]
\end{lstlisting}
The Listing \ref{haproxy_yml} defines a new docker container with the image haproxy.
The container will expose the correct ports and the custom configuration file "myHaproxy.cfg" will be mounted into the file system.


\subsection{The Database service}\label{sub:DB}
A MariaDB RDBMS cluster is configured by defining 'yml' files in the Master/Slave configuration \cite{bartholomew2013getting,wood2019mariadb}.
In a "yml" file it is in fact possible to define more containers by adding more elements in the "services" branch.

We have provided a single master database whose data is directly saved to the system disk.
As this is a master slave mode, it is important to enable logging on the file system so that debugging can be carried out and configuration errors can be analysed.
The Slave database containers have a "yml" definition similar to that of the Master: differences only concern the names and IP addresses.
We will now describe how the load balancer for Databases is configured.
The docker image used is mariadb/maxscale \cite{zaslavskiy2016full}.
\begin{figure}[ht]
    \centering
    \includegraphics[width=\linewidth]{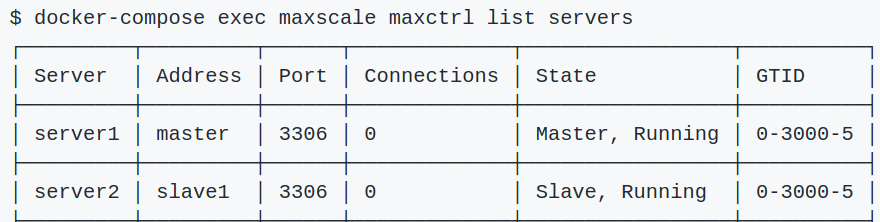}
    \caption{Database: Master and Slave}
    \label{fig:maxscale}
\end{figure}
First we need to publish the MariaDB MaxScale REST API, an HTTP interface, which generates data in JSON format, offering visual management tools.
MariaDB MaxScale splits requests in such a way that write instructions are sent to the Master container and read instructions are balanced among the Master and Slave containers.
A specific user for maxscale, with \texttt{GRANT ALL} privileges, must then be defined, acting in the Master database. 
MariaDB MaxScale is distributed with a BSL (Business Source License) and is capable of doing much more than just load balancing: it also has the ability to perform failover and switchover.
The failover mode allows to monitor and operate even if one of the nodes in the database is in an unhelpful state.
If, for example, the Master database were to crash, the Max Scale load balancer would be able to promote the Slave database to the role of new Master.
The configuration involves a few steps: definition of the servers, creation of the monitoring service, definition of the traffic routing using \texttt{readwritesplit} mode and finally the configuration of the listner using the mariadbclient protocol and its TCP port.
The resulting configuration is shown in Figure \ref{fig:maxscale}.

\subsection{Network file system}\label{sub:NFS}
The file system must be highly available, reliable and distributed.
Nodes must be able to access the file system where they will safely store data, regardless of where it is located within the infrastructure.
To do this, we have chosen to implement the GlusterFS network file system \cite{boyer2012glusterfs,selvaganesan2016insight}.
GlusterFS is distributed, open source and highly scalable.

The file system of the individual nodes has been configured to use XFS \cite{pawlowski1994nfs}.
GlusterFS includes commands for defining the list of trustworthy servers that comprise the trusted pool for sharing disk space for replication.
They are defined by executing the command 
\texttt{gluster peer probe <hostname>} command, specifying the various nodes.
In order to create the gluster volume, it is necessary to specify in sequence: the volume name, the type (e.g. Replica), and the nodes involved with their brick paths.

You need to authorise the four nodes running GlusterFS to connect to the created volumes. 
To do this, we need to specify the IP addresses for each node that we want to connect to the gluster volume.
The parameter to use is \texttt{auth.allow} and then start the volume.
Finally, the common folder where the data will be stored must be created, for example \texttt{/var/gvolume}.
This folder must be created in each node that will use it. 
To mount it, simply use the specific script \texttt{mount.glusterfs}. 
In order to speed up the start-up time of the glusterd daemon and ensure that the reboot process is automated, we recommend automount.
GlusterFS is one of the easiest persistent storage solutions to implement, combined with the use of SSD disks, which are now widely used.

\subsection{Scaling}
Scaling is simple to implement adopting our recommended design.
Our architecture enables to expand the computational power available in the system without requiring substantial structural modifications or shutting it down.
Assume, for example, that we expect to have five times as many users in November as we do the rest of the year.
In this situation, we might install more nodes, for example, by adding containers containing a Web Server instance.
The HAproxy load balancer will take care of sorting requests across nodes as described in Section\ref{sub:LB}.
Because our design involves operating in a private cloud environment, it is possible that we may need to employ more hardware and processors.
The docker setup, on the other hand, comes to our rescue.
Since all services are described by "yml" files, it will not be necessary to reconfigure the machines from scratch.
All is needed is to launch new containers within the Linux distribution and tell HAproxy which new IPs are to be used in the pool of web servers.
These procedures may be carried out without ever shutting down or disrupting the infrastructure.

\section{Conclusion and future developments}

Until recently, the term "cluster" was associated with huge corporations and data centres.
Thanks to the Open Source software available, everyone has the opportunity to deploy a Docker Cluster.
Containers allow for the expansion of infrastructure and the increase of computing capability that may be provided in a relatively short period of time.
Companies and network system engineers may model and calibrate infrastructures based on the estimated number of users by introducing additional containers and utilising the capabilities of load balancers.
Autoscaling refers to more sophisticated approaches that can generate or delete containers based on the number of users currently present.
We have achieved long-term dependability and a low RPO and RTO time using the Pilot Light model, which will ensure that we do not lose data and keep our services available to consumers in the case of various problems occur (such as hacker attacks or natural catastrophes).
In the future, we want to offer pre-configured Docker images that consumers may freely utilise.
Furthermore, we want to establish a pre-configured architecture utilising the Infrastructure as a Code (IAAC) paradigm, which allows the entire virtual structure to be described in a text file and then automated reproduced in the many organisations where it may be required.

\printbibliography
\newpage
\end{document}